# A novel Flexible and modular energy storage system for near future Energy Banks

Daniele Fargion[a], Emanuele Habib[b,*]

[a] Physics Department and INFN, Rome University Sapienza, Ple. A. Moro 2, Rome, 00185, Italy
[b] DIAEE - Astronautics, Electrical, Energy Eng. Dpt., Rome University Sapienza, via Eudossiana 18, Rome, Italy
* Corresponding author: daniele.fargion@roma1.infn.it

**Abstract**

We considered a novel energy storage system based on the compression of air through pumped water. Differently from CAES on trial, the proposed indirect compression leaves the opportunity to choose the kind of compression from adiabatic to isothermal.

The energy storage process could be both fast or slow leading to different configuration and applications. These novel storage system are modular and could be applied in different scales for different locations and applications, being very flexible in charge and discharge process. The system may offer an ideal energy buffer for wind and solar storage with no (or negligible) environment hazard.

The main features of this novel energy storage system will be showed together with overall energy and power data.

**Introduction**

Energy storage systems are of paramount interest at present, as they are mandatory in order to raise electricity production from uncontrollable renewable sources (e.g. solar energy, wind energy) [1-4]. Moreover, they can provide a better exploitation of existing power plants avoiding the construction of new power plants just to respond to growth in peak power demand, that is especially valuable in European countries where environmental impact is critical as best suitable sites has already been exploited.

There are a lot of different systems proposed for energy storage, right now the ones with commercial development are just hydrodynamic storage, for large systems, and lead based batteries for medium to small ones. Alternatives for mechanical storage are given by compressed air energy storage (CAES) that are under demonstrating operation from long time in some sites like Huntorf in Germany (operating since 1978) and McIntosh, Alabama, USA (operating since 1991). These systems are quite promising as they don't have the geographical limits of hydrodynamic storage, thus being of wider use. Nonetheless they need to provide heat to air before turbine expansion, leading to a delay in activation and fuel consumption that means it has some power generation. Moreover this requires many ancillary services for the storage plant. These systems have low energy efficiency, rated 40% - 75% [2], too, partly due to thermal issues and partly to mechanical issues related to air compression and air expansion with a variable pressure gap.

Some evolutions of CAES systems have been proposed, trying to overcome the need for heating of air before expansion. Actually, air has to be cooled after compression, too, in order to reduce its specific volume so to increase stored mass of high pressure air in the vessel. Thus, heat storage has been proposed to avoid fuel consumption for heating in, so called, adiabatic CAES [3 - 7]. This will improve energy and exergy efficiency of systems but it won't be useful to avoid activation delay that limits the kind of service CAES systems could provide to power grid.

Moreover, usual compressors and turbines are not suitable to operate with variable back pressure. So during charging phase, air is compressed up to the highest storage pressure. While before introduction in turbine, air is expanded in a valve, lowering its pressure to the lower storage pressure.

Thus, despite CAES technology has already started being exploited, a lot of improvement is possible.

In traditional CAES, compression of air takes place in the compressor, that is then moved to the storage vessel. Similarly, air is taken from the vessel and introduced in turbine for expansion. In the proposed system, air is compressed and expands directly in the storage vessel. This is done through a water piston that modifies air volume, reducing it during charge and increasing it during discharge. The water piston is used as heat storage so to absorb heat during compression and reject it during expansion, too.

The new system is thus a Hydraulic compressed air energy storage (HYCAES). It is composed of high pressure storage vessel, almost full of air when fully out of power, an atmospheric pond for water storage, a water pump and a hydraulic turbine and connecting pipes. It is not ever-new, as there are some papers illustrating similar systems [7 - 10]. In present paper, thermodynamic aspects of proposed systems will be analyzed to prove its energy feasibility.

**Polytropic transformation**

Reversible compression of air by water piston can be done through different polytropic transformations, according to heat exchange of air. Rapid compression and high volume to surface ratios provides an almost adiabatic transformation. In order to avoid limiting power to energy ratio in the system, a rapid phenomenon will be assumed, so that heat exchange through vessel is negligible. Nonetheless, a perfect mixing of water to air is assumed, so to have an almost infinite contact surface that lets any heat exchange rate be provided to air. Air and water will be assumed to the same temperature during transformation. This transformation will have the lowest possible polytropic index.

Polytropic transformation will be:

$$p_0 \cdot V_0^m = p_f \cdot V_f^m \tag{1}$$

$$T_0 \cdot V_0^{m-1} = T_f \cdot V_f^{m-1} \tag{2}$$

introducing µ=m – 1 and β=$V_0/V_f$ in (2) it becomes:

$$\frac{T_f}{T_0} = \beta^\mu \tag{3}$$

Assuming that air behaves as an ideal gas with temperature independent specific heat, while specific heat of liquid water is almost independent of transformation, energy balance for a perfectly mixed adiabatic vessel is:

$$M_a \cdot c_V \cdot (T_f - T_0) + M_w \cdot c_w \cdot (T_f - T_0) = -W \tag{4}$$

Work can be calculated straightly from polytropic equation:

$$W = \int_{V_0}^{V_f} p \cdot dV = \frac{p_0 \cdot V_0^m}{m-1}(V_0^{1-m} - V_f^{1-m}) \tag{5}$$

Mass of air is related to initial state through equation of state:

$$M_a = \frac{p_0 \cdot V_0}{R_a \cdot T_0} \tag{6}$$

Mass of water is related to its density:

$$M_w = \rho_w \cdot (V_0 - V_f) \tag{7}$$

Thus, eq. (4) becomes:

$$\frac{p_0 \cdot V_0}{R_a \cdot T_0} \cdot c_V \cdot (T_f - T_0) + \rho_w \cdot (V_0 - V_f) \cdot c_w \cdot (T_f - T_0) = -\frac{p_0 \cdot V_0^m}{m-1}(V_0^{1-m} - V_f^{1-m}) \tag{8}$$

$$\frac{c_V}{R_a} \cdot \left(\frac{T_f}{T_0} - 1\right) + \frac{\rho_w}{p_0} \cdot \frac{V_f}{V_0}\left(\frac{V_0}{V_f} - 1\right) \cdot c_w \cdot T_0 \left(\frac{T_f}{T_0} - 1\right) = \frac{1}{m-1}\left[\left(\frac{V_f}{V_0}\right)^{1-m} - 1\right] \tag{9}$$

$$\left(\frac{T_f}{T_0}-1\right)\cdot\left(\frac{c_V}{R_a}+\frac{\rho_w\cdot c_w}{p_0}\cdot\frac{T_0}{\beta}(\beta-1)\right)=\frac{1}{\mu}(\beta^\mu-1) \qquad (10)$$

$$\frac{T_f}{T_0}-1=\frac{1}{\mu}\cdot\frac{(\beta^\mu-1)}{\frac{c_V}{R_a}+\frac{\rho_w\cdot c_w}{p_0}\cdot\frac{T_0}{\beta}(\beta-1)} \qquad (11)$$

Substituting (3) in (11) it becomes:

$$\beta^\mu-1=\frac{1}{\mu}\cdot\frac{(\beta^\mu-1)}{\frac{c_V}{R_a}+\frac{\rho_w\cdot c_w}{p_0}\cdot\frac{T_0}{\beta}(\beta-1)} \qquad (12)$$

Then:

$$\mu=\frac{1}{\frac{c_V}{R_a}+\frac{\rho_w\cdot c_w}{p_0}\cdot\frac{T_0}{\beta}(\beta-1)} \qquad (13)$$

that, by defining air initial density from (6) becomes:

$$\mu=\frac{R_a}{c_V}\cdot\frac{1}{1+\frac{\rho_w\cdot c_w}{\rho_{a,0}\cdot c_v}\left(1-\frac{1}{\beta}\right)} \qquad (14)$$

So, recalling Mayer's relation, the polytropic index with respect to adiabatic index is so expressed:

$$m=1+\frac{k-1}{1+C\cdot\left(1-\frac{1}{\beta}\right)} \qquad (15)$$

where C is the ratio of heat capacity per unit volume between water and air at initial state. At low compression ratios (β → 0) polytropic is close to adiabatic. At high compression ratios it depends on C that depends on initial state on behalf. As long as initial state is almost at ambient temperature and at low pressure (same order as atmosphere), C is big around 1000. Thus even compression ratios so low as 2 lead to m values around 1.0008. This means that it is possible to reach an almost isothermal transformation through a sufficient sparkling of water during compression.

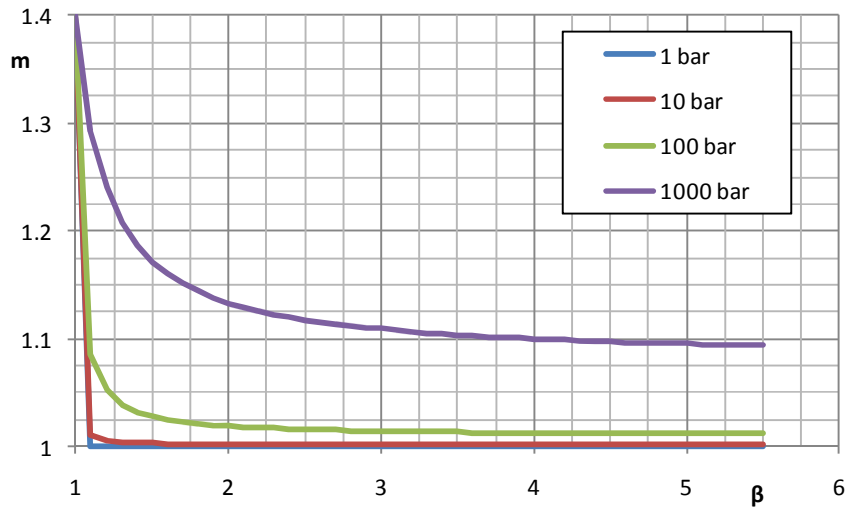

Fig. 1 - lower value of polytropic index vs. compression ratio for different initial values of air pressure

**Optimal compression ratio**

Final pressure of storage system is limited by its mechanical stability, as it should be as high as possible, as this enhances energy storage per unit mass of air. On the contrary, even though lower initial pressure means more storable energy per unit mass, it means lower initial mass, too, as stated by eq. (6).

Assuming a reversible isothermal compression during energy storage specific storable energy per unit volume is given by:

$$E = -\frac{W}{V_0} = -\frac{1}{V_0}\int_{V_0}^{V_f} p_0 \cdot V_0 \frac{dV}{V} = p_0 \cdot \ln\frac{V_0}{V_f} = p_0 \cdot \ln\frac{p_f}{p_0} \tag{16}$$

Differentiating with respect to initial pressure:

$$\frac{dE}{dp_0} = \ln\frac{p_f}{p_0} - 1 \tag{17}$$

From eq. (16) is evident that E tends to 0 both as $p_0$ tends to 0 and as $p_0$ tends to $p_f$. Thus specific storable energy per unit volume has a maximum in:

$$p_0 = \frac{p_f}{e} \tag{18}$$

This means that optimal compression ratio β is e.

This value is higher than pressure rate between initial and final pressure in operating CAES [11], even though this optimal value is not relevant for them.

**Economic issues**

CAES has been proved to be cost effective with efficiency comparable to hydrodynamic storage. It should be highlighted that in the proposed system the fully charged storage vessel is mainly filled of water rather than air. As energy is stored in air compression, this means that energy storage cost per unit volume is higher than in conventional CAES.

Nonetheless, in existing systems, compressed air has to be cooled prior to being stored, losing a lot of energy, and it has to be reheated after then, before expansion in turbine. These lead to a large amount of thermal energy loss in conventional CAES.

Conventional compressed air storage are almost isothermal, as air is cooled after compression and heated before expansion. Actually, temperature increases during storage and decreases during air extraction, thus reducing energy efficiency, but for present calculations this will be neglected. Stored energy per unit volume is thus given by the difference in air mass as specific internal energy is almost constant. Thus, neglecting the dependence of specific heat on temperature (limited to 7% in the relevant thermodynamic states of air), it could be expressed as:

$$E = (M_f - M_{0'}) \cdot c_V \cdot \frac{T_{0'}}{V} = \left(\frac{M_f \cdot T_f}{V} - \frac{M_{0'} \cdot T_{0'}}{V}\right) \cdot c_V = \left(\frac{p_f}{R} - \frac{p_{0'}}{R}\right) c_V = \frac{1}{k-1} \cdot (p_f - p_{0'}) \tag{19}$$

Thus, the ratio between (16) and (19) is vessel usage ratio:

$$\varepsilon = \frac{p_0 \cdot \ln\frac{p_f}{p_0}}{\frac{1}{k-1} \cdot (p_f - p_{0'})} \tag{20}$$

If both systems are used with the same initial and final pressure, i.e. $p_0 = p_{0'}$, usage ratio is lower than 25% for pressure ratios higher than 2.4, as shown in fig. 2.

On the other hand, optimization of such system is rather different, thus comparison of the vessel usage ratio of the proposed system with optimal pressure ratio to conventional is more meaningful. Introducing the optimal pressure ratio for proposed energy storage system, eq. (20) becomes:

$$\varepsilon = (k-1) \cdot \frac{p_f/e}{p_f - p_{0'}} = \frac{k-1}{e} \cdot \frac{1}{1 - \frac{p_{0'}}{p_f}} \tag{21}$$

Thus, for low pressure ratios usage ratio is higher as shown in fig. 2. It should be noted that Huntorf plant has a 1.57 compression ratio that corresponds to a 40.5% usage ratio. Thermodynamic analysis for Huntorf plant, based on available data [11], with unit efficiency compressor, shows that only 31% of compression work is actually gathered in energy storage, as the rest is rejected during cooling of compressed air. The proposed system doesn't need any cooling prior to storage.

Thus, the new system proposed will have a vessel cost about 2.5 times conventional CAES, but it won't have any plant cost for cooling heat exchangers or heating systems with related fuel ancillary services. So, the investment cost can be estimated to be in between of 1 - 1.5 times conventional CAES.

During operation, newly proposed HYCAES will be 3.3 times more efficient in energy storage for thermal issues only, thus break-even will be obtained in a half of the time needed for conventional CAES.

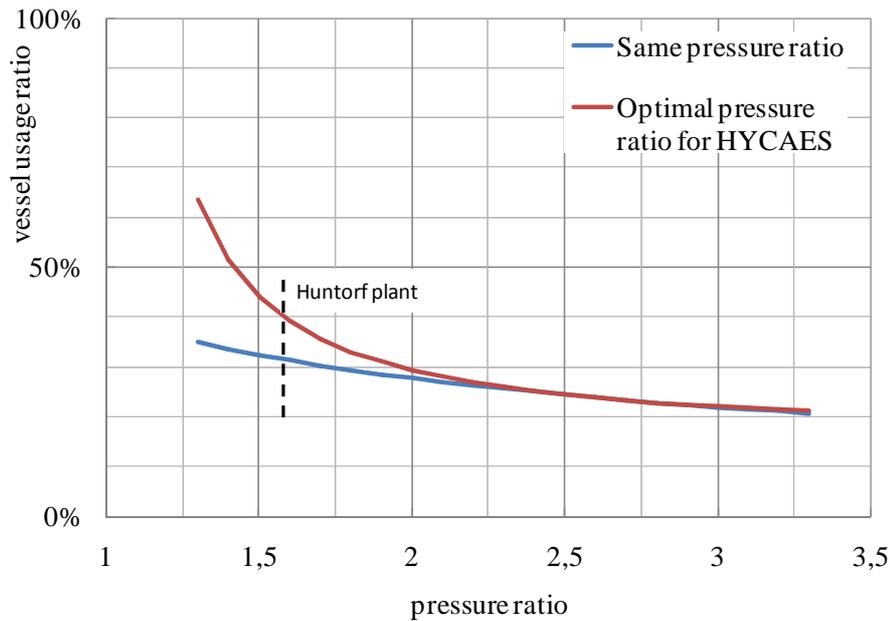

Fig. 2 - usage ratio of storage vessel of proposed system related to conventional CAES at same pressure ratio

**Conclusions**

Thermodynamic analysis of proposed system has shown that isothermal compression of air through a water piston is possible.

The proposed HYCAES is suitable for energy storage with the main advantage of no fuel nor heat storage system. Cost analysis has shown that loss in vessel usage due to water piston displacement is well compensated by the reduction of thermal energy loss after compression and of heat demand before expansion. The proposed system is thus an alternative to available large scale energy storage.

In conclusion the energy buffer is based on the combination of the well known huge thermal buffer due to the heat capacity of water over the air. This act as a thermodynamic reserve that avoids most of the energy dispersion of common CAES.

**Nomenclature**

| | | |
|---|---|---|
| c | [J/kg K] | specific heat |
| $c_v$ | [J/kg K] | specific heat for constant volume transformation for air |
| $c_p$ | [J/kg K] | specific heat for constant pressure transformation for air |
| C | $= \dfrac{\rho_w \cdot c_w \cdot R_a \cdot T_0}{p_0 \cdot c_v}$ | capacity per unit volume ratio between water and air at initial state |

| E  | [J/m³]   | storable energy per unit volume |
| m  | [-]      | general polytropic index |
| M  | [kg]     | mass |
| p  | [Pa]     | pressure |
| t  | [-]      | vessel usage ratio |
| R  | [J/kg K] | specific gas constant |
| T  | [K]      | temperature |
| V  | [m³]     | volume |
| W  | [J]      | work |

Greek symbols

| β | $= V_0/V_f$ | compression ratio |
| ε |             | vessel usage ratio |
| μ | $= m - 1$   | modified polytropic index |
| ρ | [m³/kg]     | density |

subscripts

| a  | related to air |
| f  | final state, full vessel |
| w  | related to water |
| 0  | initial state, empty vessel |
| 0' | initial state, empty vessel, conventional CAES |

**References**

[1] Electric Energy Storage Technology Options: A White Paper Primer on Applications, Costs, and Benefits. EPRI, Palo Alto, CA, 2010. 1020676.

[2] Leading the Energy Transition, Factbook, Electricity Storage, SBC Energy Institute, Gravenhage, Netherlands, 2013

[3] E. Fertig, J. Apt, Economics of compressed air energy storage to integrate wind power: A case study in ERCOT, Energy Policy, 3 9 (2011) p. 2330–2342

[4] H.L. Ferreira, R. Garde, G. Fulli, W. Kling, J.P. Lopes, Characterisation of electrical energy storage technologies, Energy 53 (2013) p. 288–298

[5] C. Bullough, C. Gatzen, C. Jakiel, M. Koller, A. Nowi, S. Zunft, Advanced Adiabatic Compressed Air Energy Storage for the Integration of Wind Energy, Proceedings of the European Wind Energy Conference, EWEC 2004, 22-25 November 2004, London UK

[6] N. Hartmann, O. Vöhringer, C. Kruck, L. Eltrop, Simulation and analysis of different adiabatic Compressed Air Energy Storage plant configurations, Applied Energy 93 (2012) p. 541–548

[7] Y.M. Kim, J.H. Lee, S.J. Kim, D. Favrat, Potential and Evolution of Compressed Air Energy Storage: Energy and Exergy Analyses, Entropy 14 (2012) p. 1501–1521

[8] Y.M. Kim, D.G. Shin, D. Favrat, Operating characteristics of constant-pressure compressed air energy storage (CAES) system combined with pumped hydro storage based on energy and exergy analysis, Energy 36 (2011) p. 6220–6233

[9] H. Wang, L. Wang, X. Wang, E. Yao, A Novel Pumped Hydro Combined with Compressed Air Energy Storage System, Energies 6 (2013) p. 1554–1567

[10] C. Qin, E. Loth, Liquid piston compression efficiency with droplet heat transfer, Applied Energy 114 (2014) p. 539–550

[11] M. Raju, S.K. Khaitan, Modeling and simulation of compressed air storage in caverns: A case study of the Huntorf plant, Applied Energy 89 (2012) p. 474–481